# Giant THz surface plasmon polariton induced by high-index dielectric metasurface

Shuai Lin[1], Khagendra Bhattarai[2], Jiangfeng Zhou[2], Diyar Talbayev[1,*]

[1]Department of Physics and Engineering Physics, Tulane University, New Orleans, Louisiana 70118, USA
[2]Department of Physics, The University of South Florida, Tampa, Florida 33620-7100, USA
*corresponding author: dtalbaye@tulane.edu

**Abstract**

We use computational approaches to explore the role of a high-refractive-index dielectric $TiO_2$ grating with deep subwavelength thickness on InSb as a tunable coupler for THz surface plasmons. We find a series of resonances as the grating couples a normally-incident THz wave to standing surface plasmon waves on both thin and thick InSb layers. In a marked contrast with previously-explored metallic gratings, we observe the emergence of a much stronger additional resonance. The mechanism of this giant plasmonic resonance is well interpreted by the dispersion of surface plasmon excited in the air\\$TiO_2$\\InSb trilayer system. We demonstrate that both the frequency and the intensity of the giant resonance can be tuned by varying dielectric grating parameters, providing more flexible tunability than metallic gratings. The phase and amplitude of the normally-incident THz wave are spatially modulated by the dielectric grating to optimize the surface plasmon excitation. The giant surface plasmon resonance gives rise to strong enhancement of the electric field above the grating structure, which can be useful in sensing and spectroscopy applications.

**Introduction**

Terahertz (THz) frequency range of electromagnetic spectrum has attracted significant fundamental research interest and technological development due to various potential applications, such as spectroscopy[1,2], bio sensing[3,4], high speed communication[5], and subwavelength imaging[6,7]. The extraordinary properties of surface plasmon polaritons (SPPs), such as strong electromagnetic field confinement, make them an important topic in THz science and technology[8,9]. THz frequency SPPs can exist on two main classes of surfaces: one is a surface of a semiconductor with a low electron or hole density that brings the bulk plasma frequency down to THz range[10,11]; the other is a microscopically corrugated or perforated metal surface on which spoof THz SPPs can occur[12]. Semiconductor indium antimonide (InSb) is a perfect candidate for the study of THz SPPs[13] because of its high electron mobility and intrinsic electron density at room temperature of $10^{16}$ cm$^{-3}$, which corresponds to about 1.8 THz bulk plasma frequency. The bulk electron density and plasma frequency are easily tunable by temperature[14,15]. In recent years, THz SPPs on InSb have been the focus of research that addressed both fundamental[16] and applied plasmonics questions, including promising sensing and spectroscopy uses[17,18]. The propagation and properties of spoof THz SPPs are determined entirely by the microscopic structure of the supporting metallic surface and cannot be easily tuned. By contrast, THz SPPs on InSb offer tunability by both temperature and very moderate magnetic field[19–21].

Investigations and utility of THz SPPs rely on a means to excite and detect them via coupling to a free-space electromagnetic wave. Various coupling schemes have been proposed, such as prism coupling[22], grating coupling[23], or a knife edge[24], among others[25]. Spoof SPPs on periodically structured surfaces can get excited directly by an incident free space wave, as the periodicity of the structure acts to select the SPP wave vector[26–30]. In this work, we explore the role of a dielectric grating as a tunable coupler between a normally-incident free-space wave and THz SPPs on thin and thick InSb layers. In a previous study, we established micrometer-thin InSb layers with metallic gratings as a platform for THz plasmonic devices; the metal grating couples the free-space wave to SPPs[17]. Most metals, including gold, silver, and copper, can be well approximated by a perfect conductor in THz range. Metallic grating offers little tunability in terms of the coupling between SPPs and free-space waves. A dielectric grating allows spatial modulation of both the phase and amplitude of the incident THz wave to manipulate the excitation of SPPs. By varying the thickness, period, and refractive index of the grating, we are able to control both the strength and frequency of the SPP resonance in THz transmission and reflection of our structure. We compute the transmission and reflection spectra of InSb layers with a dielectric grating and find that in addition to the SPP resonances present in metallic grating/InSb structures, dielectric grating/InSb structures exhibit another much stronger resonance in both transmission and reflection. This new resonance is also a SPP mode that is sustained even on thick InSb layers where the other SPP modes are no longer observable. The dispersion of the strong new resonance is well interpreted by the SPP dispersion on air/TiO$_2$/InSb trilayer structure. The dielectric grating structure exhibits strong surface electric field enhancement typical of SPPs.

Figure 1 shows the geometric configuration of our THz plasmonic structure, which consists of a grating on InSb layer. The grating has a period $d$=60 μm with equal alternating 30-μm-wide grating strips and gaps that are periodic in the $x$ direction. The $x$-polarized THz wave is incident vertically from the top of the structure. The grating period $d$ sets the wavevector $\beta_0 = 2\pi/d$ of the excited THz SPPs. The thickness and period of the dielectric grating are variable parameters that allow the tuning of the coupling to THz SPPs. While the THz refractive index of the dielectric material is also a tuning parameter, we choose titanium dioxide (TiO$_2$) as the high-refractive-index grating material and keep the refractive index fixed at $n$=9.5[31]. We vary grating thickness in the 0.2-2.4 μm range and we study the transmission and reflection spectra of this structure on 2 and 5 μm thin InSb layers and a 500 μm thick InSb wafer. The geometric parameters of this structure allow fabrication and processing using established photolithography and microfabrication methods.

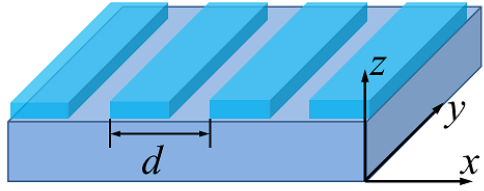

Figure 1. **Schematic of a thin InSb slab with dielectric grating.** The incident THz wave is in negative z direction with x polarization. The structure stretches indefinitely in the x and y direction. The thickness of the grating and InSb substrate varies for different simulations.

**Results**

Figure 2(a,b) shows the transmission spectrum of our plasmonic structure with InSb layer thicknesses of 2 and 5 µm. The thickness of the dielectric grating is 1 µm. The grey line is the transmission spectrum of this structure with no electrons in InSb or the bulk plasma frequency $\omega_p = 0$. The solid color lines correspond to the transmission spectra with different bulk plasma frequencies, which are labeled on the spectra. The dash color lines are the transmission spectra of the same InSb layer with a metallic grating of the same period and 200 nm thickness[17]. The transmission spectra display an overall band-pass shape. The high-pass part of transmission results from absorption and reflection of low-frequency light by conduction electrons described by the Drude model. The low-pass part of transmission results from the Fabry-Perot effect in the micrometer-thin InSb slab, and the diminished high-frequency transmission is due to the onset of the first Fabry-Perot minimum. Thus, the band-pass transmission peak is broader for the thinner InSb slab, in agreement with the Fabry-Perot description[17].

Superimposed on the band-pass shape of the transmission spectra, we find two clear high-frequency sharp transmission dips in both 5 µm and 2 µm InSb which are almost the same frequency and strength as the metallic grating spectra (dash lines), Fig. 2(a,b). These two resonance modes correspond to the odd and even SPP modes on air/InSb/air trilayers and we have studied them in detail in a previous article[17]. In addition to the high-frequency SPP resonances on thin InSb layers, a much stronger low-frequency absorption resonance is observed in the dielectric grating transmission spectra that is not present in the spectra of the metallic grating on InSb. This giant absorption resonance shows strong dependence on the bulk plasma frequency, which is indicative of its SPP origin. (We use the term *giant* to highlight the strength of this new SPP resonance and contrast it with the much weaker high-frequency resonances that are excited in the structure with both metallic and dielectric gratings.) In the following, we investigate the detailed properties of the giant low-frequency SPP resonant absorption.

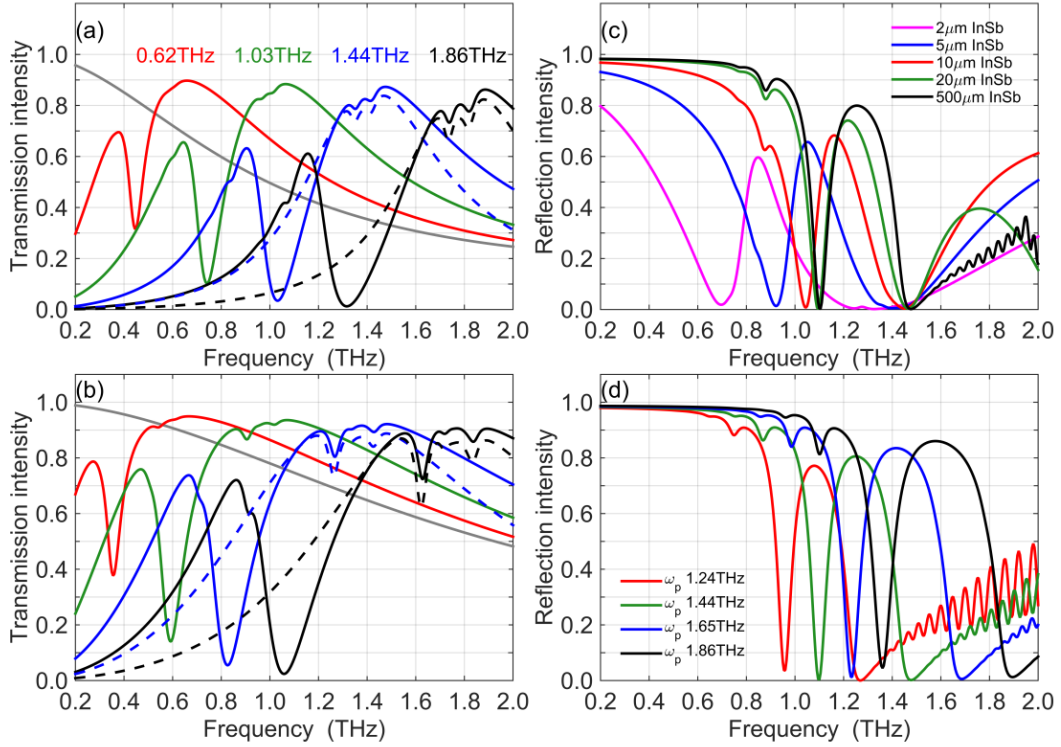

Figure 2. **Transmission and reflectance of dielectric grating/InSb structures.** (a) Transmission of 5 µm-thick InSb layers with different bulk plasma frequency (solid lines). The grey solid line is the transmission of the structure with plasma frequency $\omega_p = 0$. (b) Transmission of 2 µm-thick InSb layers with different bulk plasma frequency (solid lines). The grey solid line is the transmission of the structure with $\omega_p = 0$. Dashed lines in graph (a,b) show the transmission of structures with metallic gratings. (c) TiO$_2$ grating on InSb layers of different thickness. The bulk plasma frequency of InSb is $\omega_p = 1.44$ THz. (d) TiO$_2$ grating on 500 µm-thick InSb layer with different bulk plasma frequency. The thickness of the dielectric grating is 1 µm and the period is *d*=60 µm in all graphs.

The weak high-frequency SPP resonances vanish very quickly as the InSb slab thickness increases, Fig. 2(a,b). The transmission of the InSb structure vanishes as well due to Drude absorption and reflection by conduction electrons, so we need to compute reflectance in order to explore the behavior of the giant SPP resonance with InSb thickness. As Figure 2(c) shows, the giant SPP resonance is clear and strong in the reflection spectra of the dielectric grating on thin and thick InSb layers. The frequency of the giant resonance increases as the InSb layer thickness grows from 2 µm to 20 µm and then stays stable as the InSb thickness grows from 20 µm to 500 µm. The weak high-frequency SPP modes disappear when InSb thickness becomes larger than about 10 µm. This is because the two high-frequency modes result from mixing between SPP resonances on the upper and lower air/InSb interfaces, and this mixed mode on a thin InSb layer can sustain much larger wave vector *k* than a single propagating mode on the air/InSb interface. When InSb layer becomes sufficiently thick, the SPP modes on the upper and lower InSb surfaces become decoupled and cease to exist at high wave vectors. Effectively, the SPP loss becomes much higher for thick InSb layers and the high-frequency modes cannot get excited. In this context, the thickness of about 10 μm separates thin and thick InSb layers. For thick InSb layers, the SPP modes on the top and bottom surfaces of InSb can be considered as decoupled non-interacting waves. Figure 2(d) shows how the giant SPP resonance observed in reflectance depends on the bulk plasma frequency of InSb.

The existence of the giant SPP resonance on thick InSb layers (there is no limit on the possible InSb thickness) is one of the main findings in our work, in contrast to the vanishing of the weak high-frequency SPP resonances on InSb layers thicker than 10 μm. The robustness of the giant SPP resonance means that plasmonic devices can be fabricated on thick InSb wafers, which allows us to avoid the difficulty of manufacturing suspended grating/InSb structures with very thin InSb layers required for metallic gratings.

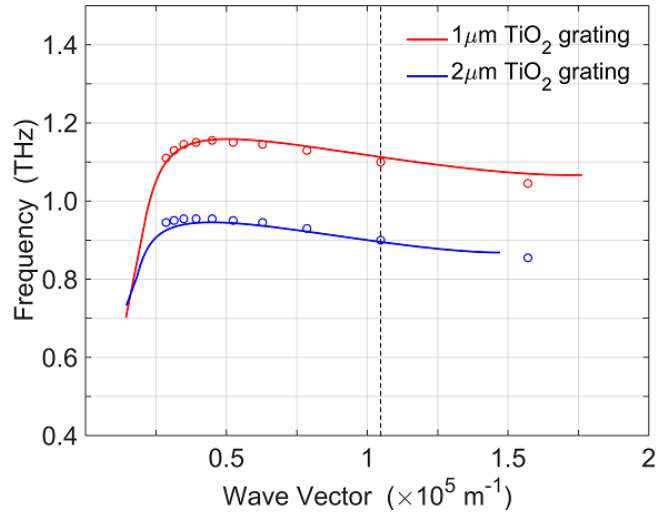

Figure 3. **Computational (open circles) vs. theoretical (solid lines) dispersion of the giant SPP resonance.** The bulk plasma frequency of InSb is $\omega_p = 1.44$ THz. The InSb layer thickness is 500 μm. The vertical dash line shows the wave vector for $d = 60$ μm grating.

We now demonstrate that the behavior of the giant SPP resonance in the reflection spectra agrees well with SPP theory of air/TiO$_2$/InSb trilayer structure in which air and InSb have infinite thicknesses. We calculate the theoretical dispersion relation of this trilayer from[32]

$$e^{-4k_1 a} = \frac{k_1/\epsilon_1 + k_2/\epsilon_2}{k_1/\epsilon_1 - k_2/\epsilon_2} \frac{k_1/\epsilon_1 + k_3/\epsilon_3}{k_1/\epsilon_1 - k_3/\epsilon_3} \quad (1)$$

and

$$k_i^2 = \beta^2 - k_0^2 \epsilon_i, \quad (2)$$

where $i=1,2,3$, $k_0 = \omega/c$ is the wavevector of THz wave in vacuum, $a$ is the thickness of the TiO$_2$ layer, $\beta$ is the SPP wavevector along the air/TiO$_2$/InSb interface, $\epsilon_1$, $\epsilon_2$ and $\epsilon_3$ are the dielectric permittivities of TiO$_2$, InSb and air, respectively. The quantities $k_1$, $k_2$, and $k_3$ are the imaginary wave vectors for TiO$_2$, InSb, and air in the direction perpendicular to the interface. The complex permittivity of InSb is described by the Drude model:

$$\epsilon_1(\omega) = \epsilon_\infty \left(1 - \frac{\omega_p^2}{\omega^2 + i\omega\gamma}\right), \tag{3}$$

where $\epsilon_\infty$ is the background dielectric constant, which is 15.6 for InSb, $\omega_p$ is the bulk plasma frequency, $\gamma = 0.3 \times 10^{12}$ THz is the scattering rate for InSb at low temperature[33].

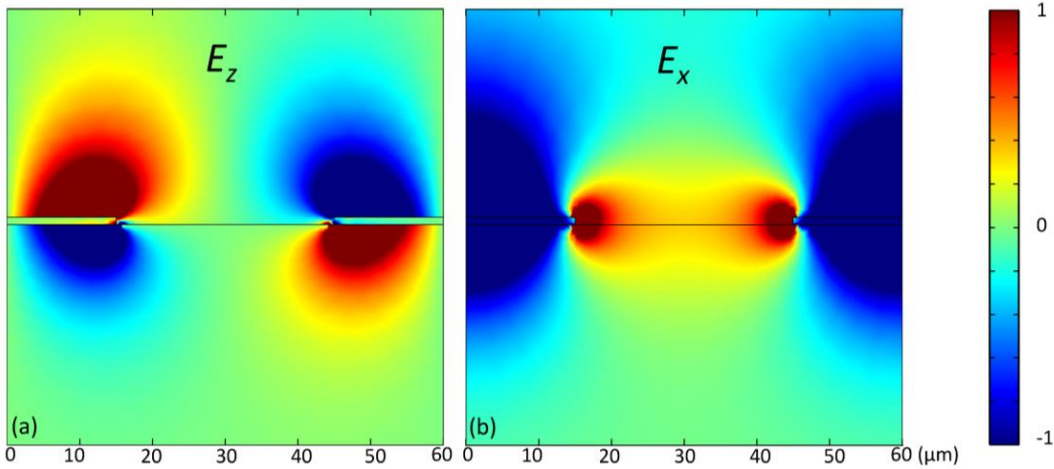

Figure 4. **Spatial distribution of electric field at the frequency of the giant SPP resonance.** (a) Distribution of the $z$ component of the electric field in the standing SPP wave in a 60x60 μm square around a grating strip representing one grating period. The TiO$_2$ strips are 1 μm thick and cover 15-μm lengths on each side of the panel for a total strip width of 30 μm. (b) Distribution of the $x$ component of the total electric field. InSb bulk plasma frequency in both graphs is $\omega_p = 1.44$ THz. InSb thickness is 500 μm. The color scale is the same for both graphs and uses arbitrary units.

Figure 3 compares the theoretical trilayer SPP dispersion (solid lines) calculated from equations (1)-(2) with the dispersion (open circles) extracted from the simulated reflection spectra of the grating/InSb structure. The bulk plasma frequency is 1.44 THz in Figure 3. To obtain the dispersion relation from our numerical reflection simulations, we need to identify the wavevector of the standing SPP mode induced by the grating. By inspecting $z$ component of the electric field at the resonance frequency, we find that the giant SPP resonance in the reflection spectra, Fig. 2 (d), corresponds to the standing wave mode with wave vector $\beta_0 = 2\pi/d$, where $d$ is the grating period. We extract the SPP dispersion from the simulated reflection spectra by varying the grating period $d$. This dispersion matches very well the theoretical dispersion calculated from the air/TiO$_2$/InSb trilayer structure, Eqs. (1)-(2). The trilayer theory confirms our assignment of the giant absorption resonance as a SPP mode. It also demonstrates the tunability of the absorption resonance frequency via the dielectric grating period.

Why does the theory of an SPP resonance on the continuous air/TiO$_2$/InSb trilayer describe so well the dispersion of an SPP resonance found on the TiO$_2$ grating/InSb structure? The answer is provided by Fig. 4(a) that shows the distribution of the $z$ component of the electric field at the SPP resonance frequency. The field $E_z$ is the hallmark of the SPP resonance, e.g., this field is zero at off-resonance frequencies. The electric field $E_z$ is concentrated above and below the TiO$_2$ strips of the grating, as shown in Fig. 4(a). While the grating is necessary to excite the giant SPP resonance, the presence of the air gaps does not appreciably perturb the dynamics of the standing SPP wave, which is why it closely resembles the dynamics of the propagating SPP wave on a continuous air/TiO$_2$/InSb structure. We tentatively ascribe the weaker resonances in the reflection spectra found below the frequency of the giant SPP resonance, Fig. 2 (d), to higher order SPP modes with wavevectors that are integer multiples of $\beta_0$. As shown by the shape of the dispersion curve in Fig. 3, the derivative $d\omega/dk$ is negative for high wave vectors; qualitatively, higher order modes of larger $k$ are expected to have a lower frequency than the fundamental SPP mode. In the discussion below, we focus only on the fundamental mode.

**Discussion**

A major difference between the structure with the dielectric grating and a similar structure with a metallic (gold) grating is the presence and unique properties of the giant SPP resonance. Figure 5(a) shows that the resonance strength increases quickly as the dielectric grating thickness increases from 0.2 μm to 1.4 μm; the resonance strength remains unchanged until the grating thickness reaches about 2.2 μm. The grating couples the incident free-space wave to the standing SPP modes by modulating the spatial distribution of the THz electric field along the InSb surface. To achieve better coupling, the modulated electric field should mimic as close as possible the electric field of the standing SPP wave. The gold grating modulates the THz field by imposing perfect-metal boundary conditions periodically along the InSb surface (gold is well approximated by the perfect metal at THz frequencies)[17]. An extended dielectric layer atop InSb allows a control of amplitude and phase of the THz field underneath the InSb surface. The high refractive index of TiO$_2$ ($n$=9.5) is the main reason why such a thin (~ 1 μm) dielectric layer efficiently modulates both the amplitude and phase of THz field. In the simplest optical model, the THz amplitude and phase underneath the TiO$_2$ layer inside InSb can be computed from Fresnel reflection coefficients and the dielectric layer thickness (the optical path of the THz wave) and compared to the amplitude and phase of the wave incident on bare InSb. Figure 5(b) shows how both amplitude and phase of THz field $E_x$ inside InSb at the SPP resonance frequency change with the dielectric layer thickness; the THz phase and amplitude are compared to the phase and amplitude of the same wave in the absence of any dielectric, which corresponds to the gaps of the dielectric grating. Figure 4(b) shows the spatial distribution of electric field $E_x$ in the standing SPP wave. The $E_x$ oscillation is characterized by a phase difference of π between the gaps and strips of the dielectric grating; the oscillation is completely out of phase between these two regions. Therefore, we can expect strongest grating coupling efficiency when the phase difference between gaps and strips for the incident $E_x$ inside InSb is close to π, the maximum phase contrast. This phase contrast approaches 0.83π at around 2.5 μm grating thickness and the amplitude contrast reaches it maximum at 1.5 μm grating thickness, Fig. 5(b). The combined thickness dependence of the amplitude and phase contrast at low grating thicknesses largely explains the fast rise of the coupling efficiency from 0.2 μm to 2 μm, Fig. 5(a).

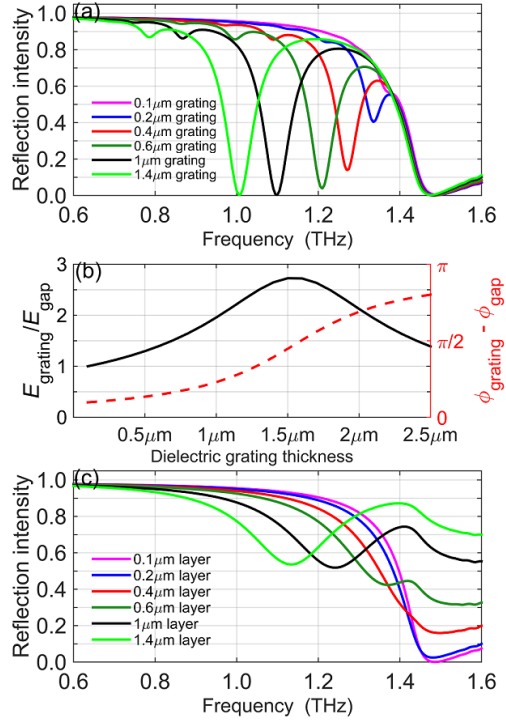

Figure 5. **Dependence of optical properties on the dielectric grating thickness.** (a) Reflectance of dielectric grating/InSb structures with different grating thickness. InSb thickness is 500 μm. (b) Intensity and phase contrast of electric field just underneath InSb surface between the grating strips and air gaps. The frequency is chosen at the resonance frequency of 1 μm grating, 1.1 THz. (c) Reflectance of continuous $TiO_2$ layers on InSb. A broad anti-reflection feature develops with increasing thickness. InSb bulk plasma frequency in all graphs is $\omega_p = 1.44$ THz.

In addition to the intensity changes, the resonance frequency also becomes lower as the thickness of grating increases. Such frequency shift can be well explained by the dispersion of air/$TiO_2$/InSb trilayer system as shown in Fig. 3, where the SPP resonance frequency decreases as the thickness of $TiO_2$ layer increases. Another remarkable feature is that the thickness of $TiO_2$ grating is much smaller than the wavelength of incident wave. For instance, the resonance wavelength for 1.4 μm grating is $\lambda = 300$ μm, resulting in the wavelength-to-thickness ratio $\lambda_{TiO_2}/t \approx 23$, where $\lambda_{TiO_2} = \lambda/9.5$ is the wavelength inside the $TiO_2$ medium. Our grating is extremely thin in contrast to dielectric gratings in other applications[34], where the wavelength-to-thickness ratio, $\lambda_{grating}/t \sim 1$, in order to create the necessary phase retardance for a wave propagating through the grating layer. The ultra-thin grating greatly relieves the challenges for film deposition and the patterning process in the fabrication.

The remarkable feature of the giant SPP resonance is its strength relative to the strength of the weaker high-frequency SPP modes that are also present in the metallic grating sensor, Fig. 2(a,b). This strength is derived partly from the very efficient coupling between the free space and the SPP waves, as described in the previous paragraph. Another contribution to the SPP resonance strength is illustrated in Fig. 5(c), where we plot the reflectance of continuous TiO$_2$ layers of different thickness on InSb. Below the plasma reflectance edge, we observe a broad feature with a thickness-dependent center frequency. This feature results from the anti-reflection-coating effect of the continuous TiO$_2$ film. The energy that is not reflected is absorbed inside InSb by conduction electrons. While this broad anti-reflection feature is not evident in the reflection spectra of the grating/InSb structure, Fig. 5(a), it likely plays a role in enhancing the giant SPP resonance by allowing a larger fraction of the incident energy to couple and be absorbed as the SPP oscillation.

The excitation of SPPs is typically accompanied by an enhancement of the total electric field intensity at the resonance frequency. In our structure, the amplitude of $E_x$ can be taken as an indicator of the excited SPP strength and the surface electric field enhancement. In the simulated reflection response with 500 μm InSb, we determine the amplitude of the *x* component of the electric field above the dielectric grating strips at the SPP resonance frequency. We compare the $E_x$ amplitude for dielectric gratings of different thickness on 500 μm InSb and for a continuous dielectric layer with the same thickness on 500 μm InSb. Figure 6 shows the $E_x$ amplitude ratio between the dielectric grating and the continuous dielectric layer. The surface electric field enhancement due to the excitation of SPPs increases quickly as the grating thickness grows from 0.2 μm to about 1 μm; the enhancement ratio decreases slowly above the 1.5 μm thickness, Fig. 6. Thus, this trend agrees well with the phase and intensity modulation in Fig. 5(b). The dielectric grating achieves around 4 times electric field enhancement compared with the bare dielectric layer without grating structure. We expect that higher enhancement ratios will be possible via the optimization of other grating parameters, such as refractive index, period, and gap/strip width ratio. Surface-plasmon enhanced electric field generally leads to stronger light-matter interaction, which can be useful for spectroscopy and sensing applications[35–37].

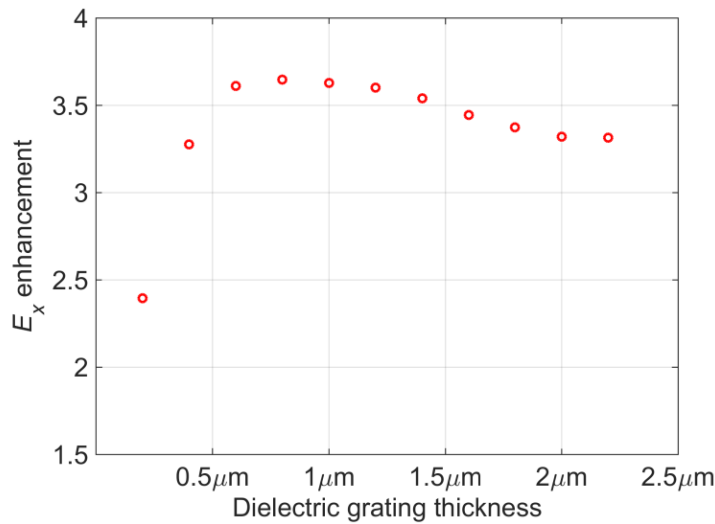

Figure 6. **Electric field enhancement due to the giant SPP resonance.** Enhancement is plotted as the ratio of the *x* component $E_x$ between the TiO$_2$ grating and an equal-thickness continuous TiO$_2$ layer on 500 μm InSb. The grating period is *d*=60 μm. The bulk plasma frequency is $\omega_p = 1.44$ THz.

To summarize, we investigated the properties of a tunable dielectric grating coupler of a free-space THz wave to standing SPP modes on thin and thick InSb layers. We found a giant SPP resonance in transmission and reflection spectra of this plasmonic structure and successfully described its dispersion. The frequency and strength of the giant resonance can be manipulated by varying the geometric parameters of the grating, such as period and thickness. The refractive index of the grating material offers another tuning variable. The dielectric grating/InSb structure achieves about 4 times surface field enhancement over the continuous dielectric layer on InSb. The normal-incidence transmission or reflection mode operation of the plasmonic structure and the photolithography-friendly fabrication make our structure a flexible and competitive platform in THz plasmonic applications.

**Methods**

This dielectric grating/InSb structure is modeled by a commercial finite element simulation platform COMSOL Multiphysics. This work is simulated in the frequency domain of COMSOL RF module. Considering the complexity of this structure and to maximize the computation speed, the simulation is modeled in 2D by setting the structure infinite and translationally invariant in the $y$ direction. In simulation, we set both the superstrate and substrate as air (or vacuum) for convenience. The use periodic boundary conditions in our simulation.

**Acknowledgements**

The work at Tulane was supported by NSF Award No. DMR-1554866. The work at USF was supported by the Alfred P. Sloan Research Fellow grant BR2013-123 and by KRISS grant GP2016-034.

14. Oszwaldowski, M. & Zimpel, M. Temperature dependence of intrinsic carrier concentration and density of states effective mass of heavy holes in InSb. *J. Phys. Chem. Solids* **49,** 1179–1185 (1988).

15. *Handbook series on semiconductor parameters. Vol. 2: Ternary and quaternary III-V compounds*. (World Scientific, 1999).

16. Stepanenko, O. *et al.* Compact mid-IR isolator using nonreciprocal magnetoplasmonic InSb mirror. in *2016 41st International Conference on Infrared, Millimeter, and Terahertz waves (IRMMW-THz)* 1–2 (2016). doi:10.1109/IRMMW-THz.2016.7758556

17. Lin, S., Bhattarai, K., Zhou, J. & Talbayev, D. Thin InSb layers with metallic gratings: a novel platform for spectrally-selective THz plasmonic sensing. *Opt. Express* **24,** 19448–19457 (2016).

18. Liu, H. *et al.* Tunable Terahertz Plasmonic Perfect Absorber Based on T-Shaped InSb Array. *Plasmonics* **11,** 411–417 (2016).

19. Hu, B., Zhang, Y. & Wang, Q. J. Surface magneto plasmons and their applications in the infrared frequencies. *Nanophotonics* **4,** 383–396 (2015).

20. Deng, L. *et al.* Direct optical tuning of the terahertz plasmonic response of InSb subwavelength gratings. *Adv. Opt. Mater.* **1,** 128–132 (2013).

21. Rivas, J. G., Janke, C., Bolivar, P. H. & Kurz, H. Transmission of THz radiation through InSb gratings of subwavelength apertures. *Opt. Express* **13,** 847–859 (2005).

22. O'Hara, J. F., Averitt, R. D. & Taylor, A. J. Prism coupling to terahertz surface plasmon polaritons. *Opt. Express* **13,** 6117–6126 (2005).

23. O'Hara, J. F., Averitt, R. D. & Taylor, A. J. Terahertz surface plasmon polariton coupling on metallic gratings. *Opt. Express* **12,** 6397–6402 (2004).
12